\documentclass{article}
\usepackage{graphicx}

\usepackage{arxiv}

\usepackage[utf8]{inputenc} % allow utf-8 input
\usepackage[T1]{fontenc}    % use 8-bit T1 fonts
\usepackage{hyperref}       % hyperlinks
\usepackage{url}            % simple URL typesetting
\usepackage{booktabs}       % professional-quality tables
\usepackage{amsfonts}       % blackboard math symbols
\usepackage{nicefrac}       % compact symbols for 1/2, etc.
\usepackage{microtype}      % microtypography
\usepackage{lipsum}
\usepackage{algorithm}
\usepackage{algorithmic}
\usepackage{multirow}
\usepackage{multicol}
\title{Few Shot Domain Adaptation for \textit{in situ} Macromolecule Structural Classification in Cryo-electron Tomograms}

\author{
Liangyong Yu \\
Computational Biology Department \\
Carnegie Mellon University \\
Pittsburgh, 15213, USA \\
\And
Ran Li \\
Department of Automation \\
Tsinghua University \\
Beijing, 100084, China \\
\And
Xiangrui Zeng \\
Computational Biology Department \\
Carnegie Mellon University \\
Pittsburgh, 15213, USA \\
\And
Hongyi Wang \\
Department of Electronic Engineering \\
Tsinghua University \\
Beijing, 100084, China \\
\And
Jie Jin \\
Institute of Automation \\
Chinese Academy of Science \\
Beijing, 100190, China \\
\And
Ge Yang \\
Institute of Automation \\
Chinese Academy of Science \\
Beijing, 100190, China \\
\And
Rui Jiang \\
Department of Automation \\
Tsinghua University \\
Beijing, 100084, China \\
\And
Min Xu \\
Computational Biology Department \\
Carnegie Mellon University \\
Pittsburgh, 15213, USA \\
}

\begin{document}
\maketitle

\begin{abstract}
\textbf{Motivation:} Cryo-Electron Tomography (cryo-ET) visualizes structure and spatial organization of macromolecules and their interactions with other subcellular components inside single cells in the close-to-native state at sub-molecular resolution. Such information is critical for the accurate understanding of cellular processes. However, subtomogram classification remains one of the major challenges for the systematic recognition and recovery of the macromolecule structures in cryo-ET because of imaging limits and data quantity. Recently, deep learning has significantly improved the throughput and accuracy of large-scale subtomogram classification. However often it is difficult to get enough high-quality annotated subtomogram data for supervised training due to the enormous expense of labeling. To tackle this problem, it is beneficial to utilize another already annotated dataset to assist the training process. However, due to the discrepancy of image intensity distribution between source domain and target domain, the model trained on subtomograms in source domain may perform poorly in predicting subtomogram classes in the target domain. \\ 
\textbf{Results:} In this paper, we adapt a few shot domain adaptation method for deep learning based cross-domain subtomogram classification. The essential idea of our method consists of two parts: 1) take full advantage of the distribution of plentiful unlabeled target domain data, and 2) exploit the correlation between the whole source domain dataset and few labeled target domain data. Experiments conducted on simulated and real datasets show that our method achieves significant improvement on cross domain subtomogram classification compared with baseline methods.\\
\textbf{Availability:} https://github.com/xulabs/aitom \\
\textbf{Contact:} \href{mxu1@cs.cmu.edu}{mxu1@cs.cmu.edu}\\
%\textbf{Supplementary information:} Supplementary data are available at \textit{Bioinformatics} online.
\end{abstract}

% keywords can be removed
\keywords{Few Shot Domain Adaptation \and Macromolecule Classification \and Cryo-electron Tomograms}

\section{Introduction}\label{introduction}

Plentiful complex biochemical processes and subcellular activities sustain the dynamic and complex cellular environment, in which a mass of intricate molecular ensembles participate. A comprehensive analysis of these ensembles \textit{in situ}\footnote{At their original locations.} inside single cells would play an essential role in understanding the molecular mechanisms of cells. Cryo-electron Tomography (cryo-ET), as a revolutionary imaging technique for structural biology, enables the \emph{in situ} 3D visualization of structural organization information of all subcellular components in single cells  in a close-to-native state at submolecular resolution. Thus cryo-ET can bring new molecular machinery insights of various cellular processes by systematically visualizing the structure and spatial organizations of all macromolecules and their spatial interactions with all other subcellular components in single cells at unprecedented resolution and coverage. 

In particular, because of fractionated total electron dose over entire tilt series \cite{bartesaghi2008classification}, we need to average multiple subtomograms\footnote{Subtomograms are subvolumes extracted from a tomogram, and each of them usually contains one macromolecule} that contain identical structures in order to get high SNR subtomogram average representing higher resolution of the underlying structure \cite{briggs2013structural}. However, the macromolecule structures in a cell are highly diverse. Therefore, it is necessary to first accurately classify these subtomograms into subsets of structurally identical macromolecules. This is performed by subtomogram classification. Systematic structural classification of macromolecules is a vital step for the systematic analysis of cellular macromolecular structures and functions \cite{irobalieva2016cellular} in many aspects including macromolecular structural recovery. However, such classification is very difficult, because of the structural complexity in cellular environment as well as the limit of data collection such as missing wedge effects \cite{bartesaghi2008classification}. %All of these factors produce the low fidelity of tomograms. 
Therefore, for successful automatic and systematic recognition and recovery of macromolecular structures captured by cryo-ET, it is imperative to have an efficient and accurate method for subtomogram classification.

% However, structural discrimination of macromolecules in cryo-ET images is restrained by the complicated intracellular environment, versatile conformations of assemblies, and the limitations of cryo-ET imaging. 

With the technological breakthrough of cryo-ET and the development of image acquisition automation, collecting tomograms containing millions of macromolecules is no longer the obstacle for researchers, and methods based on deep-learning have been proposed to address the issue of high-throughput subtomogram classification thanks to the high-throughput processing capability of deep learning. Different architectures of Convolutional Neural Network (CNN) have been explored \cite{che2018improved}. %, and structures of specific aggregates have been dissected through a deep-learning based classification method \cite{guo2018situ}. 
Despite the significant superiority in speed, accuracy, robustness and scalability compared to traditional methods, these supervised deep learning based subtomogram classification methods often suffer from the high demand of annotated data. Currently labeling is done by a combination of computational template search and manual inspection. However, in practise, template search is time-consuming and quality control through manual inspection is laborious. The complicated structure and distortion caused by noise make subtomogram images hard to distinguish by the naked eyes even by experts, which is a major obstacle for the manual quality insurance of the annotation. 
% Which in principle is not feasible to obtain on account of the huge cost of time and effort for labeling and the lack of knowledge of macromolecules.\\

An intuitive idea to tackle the problem of insufficient annotated data is to utilize a separated auxiliary dataset, which has abundant labeled samples, to assist subtomogram classification. Such auxiliary dataset is obtained from a separate imaging source or from simulation. Therefore the auxiliary dataset and our target dataset have the same structural classes but different image intensity distribution. The difference can be attributed to discrepant data acquisition conditions, such as different Contrast Transfer Function (CTF), signal-to-noise ratio (SNR), resolution, backgrounds, etc. The source domain is defined as the domain that the auxiliary dataset belongs to, and the target domain is defined as the domain that the evaluation dataset belongs to. In our case, we assume that we have plenty of labeled subtomograms in the source domain, but only few labeled samples in the target domain are accessible. This is due to the difficulty to annotate the data in target domain. For example, the real cryo-ET data in the target domain acquired from cryo-ET (real dataset) might be extremely time-consuming to annotate. On the other hand, we can generate simulated cryo-ET data in the source domain on the computer as the separated auxiliary dataset to assist us to improve the prediction accuracy of the real dataset in the target domain. Unfortunately, because of the image intensity distribution discrepancy between the source domain and the target domain, a deep learning model trained on the source domain perform poorly on the target domain due to dataset shift \cite{quionero2009dataset}. 

\begin{figure*}[h!]  
\centering
 \includegraphics[width=1.0\textwidth]{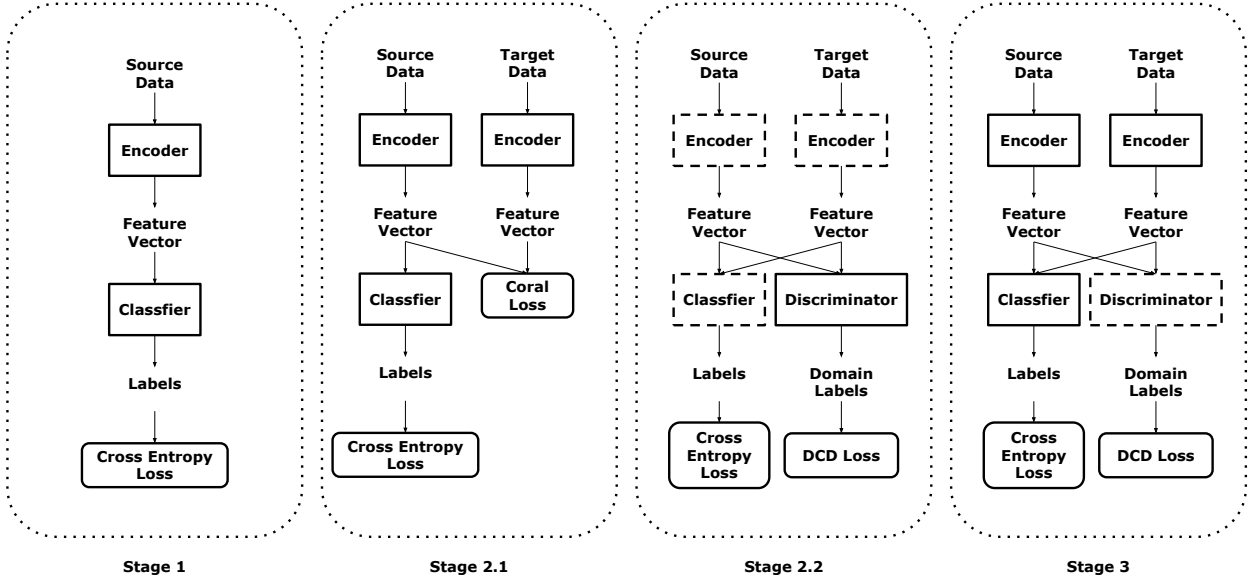}
  \caption{\label{fig:introduction} The flowchart of our method. The model whose edge is imaginary line represents that its parameters are fixed. In Stage 1, an encoder $f_{\phi}$ and a classifier $g$ are initially trained using data in source domain (Section \ref{encoderandclassifier}). In Stage 2, a discriminator $\mathcal{D}$ is trained to identify the domain of each subtomogram (Section \ref{stage2}). In Stage 3, labeled data in both domains are used to fine-tune the encoder $f_{\phi}$ with the assistance of discriminator $\mathcal{D}$ (Section \ref{stage3}).}  
\end{figure*} 

Domain Adaptation \cite{blitzer2006domain} is an effective way to solve this problem. This approach resolves the discrepancy of data distribution between source domain and target domain. One type of domain adaptation fine-tunes a trained neural network on source domain, which makes it perform well on both source domain and target domain. Another type of domain adaptation transforms target/source data in order to make it get close to the image intensity distribution of another domain [e.g.  \cite{alam2018speaker}]. Therefore, neural network doesn't need to distinguish two domains, because their image intensity distributions are similar by properly transforming the input data. Domain adaptation can also be categorized into unsupervised and supervised approaches: Unsupervised domain adaptation (UDA) requires large amount of data but doesn't need target labels [e.g. \cite{long2016unsupervised}], while supervised domain adaptation (SDA) requires target labels to be given [e.g. \cite{garcia2014supervised}]. Nowadays these two methods are the mainstream methods to reduce distribution discrepancy in source domain and target domain. However, in our cryo-ET dataset, the methods based on UDA and SDA have obvious defects: (1) UDA can't utilize the information of labeled data in target domain, therefore intra-class relationship between source domain and target domain is neglected. (2) Often, due to annotation difficulty, there are too few labeled data in target domain that SDA can't reach satisfactory results. 

Therefore, we propose a method for Few-Shot Domain Adaptation: Few-Shot Fine-Tuning domain adaptation(FSFT). Few-Shot means that each class contains only very few labels in the target domain [e.g. \cite{motiian2017few}]. Generally, for each class, we only use three to seven labels in the target domain. The flowchart of our method is presented on Figure \ref{fig:introduction}. It contains three components: encoder $f_{\phi}$, classifier $g$ and discriminator $\mathcal{D}$. Encoder $f_{\phi}$ extracts every subtomogram into a feature vector\footnote{Feature vectors represent the output of encoder $f_{\phi}$}; classifier $g$ transforms each feature vector into a one-hot label, which presents the class of each subtomogram; discriminator $\mathcal{D}$ identifies which domain the feature vectors belong to. The detailed training procedure is explained in the following section.

% We train the encoder and classifier using source data in stage 1. In the next stage, UDA based on Deep CORAL \cite{sun2016deep} is applied. Stage 3 represents the SDA following the UDA similar to Few-Shot Adversarial Domain Adaptation (FADA) \cite{motiian2017few}: first we train a discriminator to distinguish the domain of embedding space generated by encoder, and next we train our encoder and classifier again to confuse discriminator. 

We have evaluated our method on both simulated and real datasets. Compared with popular baseline methods, our method achieves significantly higher classification accuracy. Additionally, related works and result analysis are presented in supplementary document.

Our main contributions are summarized as follows: 
\begin{itemize}
    \item We are the first to use few-shot domain adaptation for cross-domain subtomogram classification.  
    \item We directly train the discriminator without adversarial training in the training procedure, comparing to FADA \cite{motiian2017few}.
    \item We introduce a mechanism of partly-shared parameters of encoder $f_{\phi}$ between source domain and target domain. The layers whose parameters are shared by two domains are called domain-independent layers, and the other layers are called domain-related layers (Section \ref{stage2_1}).
    \item We combine domain discrimination for the output of independent layers and shared layers (Section \ref{SDA}).
\end{itemize}

\section{Methods}\label{Method}

 In this section, we describe our model in details. Our training strategy contains three stages. Stage 1: an encoder $f_{\phi}$ and a classifier $g$ are initially trained using data in source domain (Section \ref{encoderandclassifier}). Stage 2: a discriminator $\mathcal{D}$ is trained to identify the domain of each subtomogram (Section \ref{stage2}). Stage 3: labeled data in both domains are used to fine-tune the encoder $f_{\phi}$ with the assistance of discriminator $\mathcal{D}$ (Section \ref{stage3}). Stage 2.1 is Unsupervised Domain Adaptation while Stage 2.2 and Stage 3 are Supervised Domain Adaptation.
\renewcommand{\algorithmicrequire}{ \textbf{Input:}} %Use Input in the format of Algorithm
\renewcommand{\algorithmicensure}{ \textbf{Output:}} %UseOutput in the format of Algorithm

 \begin{algorithm}[htb]
\caption{Overall algorithm}
\begin{algorithmic}[1]
\REQUIRE ~~\\
Encoder in source domain: $f^\textrm{0} \circ f^\textrm{s}$ \\
Encoder in target domain: $ f^\textrm{0} \circ f^\textrm{t} $ \\
Classifier $g$, discriminator $\mathcal{D}$. \\
\ENSURE ~~\\
Trained $f^\textrm{0} \circ f^\textrm{s}$, $f^\textrm{0} \circ f^\textrm{t}$, classifier $g$ and discriminator $\mathcal{D}$.\\
\STATE Train $f^\textrm{0} \circ f^\textrm{s}$ and classifier $g$ using source subtomograms (Stage 1) \\
\STATE Train $f^\textrm{0} \circ f^\textrm{s}, f^\textrm{0} \circ f^\textrm{t}$ and $g$ using unlabeled target subtomograms (Stage 2.1) by algorithm \ref{algorithm2}\\
\STATE Train discriminator $\mathcal{D}$ using labeled target and source subtomograms (Stage 2.2) \\
\STATE Fine-tune $f^\textrm{0} \circ f^\textrm{t}$ and classifier $g$ with the assistance of discriminator $\mathcal{D}$ and labeled target and source subtomograms (Stage 3) \\

\end{algorithmic}
\end{algorithm}

%  First we will utilize data in source domain to train encoder and classifier, and second we will implement Unsupervised Domain Adaptation for encoder. The last step which implements SDA is similar to adversarial network: we train a discriminator to distinguish from which domain the subtomogram samples are generated. And next the parameters of encoder and classifier are calibrated to confuse the discriminator.

\subsection{Stage 1: Initialize encoder $f_{\phi}$ and classifier $g$}\label{encoderandclassifier}
 A series of subtomogram samples in source domain $X^\textrm{s} = (x^\textrm{s}, y^\textrm{s})$ are provided in this section. We apply a 3D encoder $f_{\phi}$, which maps each subtomogram into a feature vector in embedding space. We introduce an embedding function $f_{\phi}(\cdot)$ to represent the encoder $f_{\phi}$. Because the parameters of encoder $f_{\phi}$ are partly shared between source domain and target domain, the embedding function can be composited by two parts: the domain-related function $f^{\textrm{t}}(\cdot)$ or $f^{\textrm{s}}(\cdot)$, and the domain-independent function $f^{\textrm{0}}(\cdot)$. That's to say, we apply $f^{\textrm{0}} \circ f^{\textrm{s}}(\cdot)$ for source domain and $f^{\textrm{0}} \circ f^{\textrm{t}}(\cdot)$ for target domain. 
 
The application of the partly-shared encoder $f_{\phi}$ is based on the assumption that different domains have similar high level feature (including details), because the structure of subtomograms in the same class but from different domains are similar; but their low level features are different such as edges due to image intensity difference between domains. The front part is more for low level features and the back part for high level features. In other words, the front parts of encoder $f^0$ of $f_{\phi}$ extract the common structural features of both domains and remove the domain-related features such as image parameters and SNR. The back parts $f^\textrm{s}$ and $f^\textrm{t}$ further extract their common feature into embedding space. Second, a classifier $g$ maps feature vectors into one-hot labels, which is represented by a prediction function $g(\cdot)$. 
 
 We update the encoders in source domain:$f^\textrm{0} \circ f^\textrm{s}$ and encoders in target domain:$f^\textrm{0} \circ f^\textrm{t}$ and classifier $g$ by the following equation:
 \begin{equation}\label{equation1}
    \theta \leftarrow \theta - \frac{1}{n}\beta \nabla_{\theta} [-\mathop{\sum}_{i=1}^{n}y_{i}^{s}\log(g\circ f^{\textrm{0}} \circ f^{\textrm{s}}(x_{i}^{s}))]
\end{equation}

The loss function is:
\begin{equation}\label{equation1_1}
    L^\textrm{C} = -\mathop{\sum}_{i=1}^{n}y_{i}^{s}\log(g\circ f^{\textrm{0}} \circ f^{\textrm{s}}(x_{i}^{s}))
\end{equation}

We set $n$ as batch size. $x_{i}^{s}$ represents the i-th subtomogram image, and $y_{i}^{s}$ represents the i-th subtomogram label in each subtomogram sample batch.

% \begin{equation}\label{equ:loss1}
%     L=\sum{-\log s_i}.
%     % p_\phi(y=k|x)=\frac{exp(-d(f_\phi(x),c_k))}{\sum_{k'}exp(-d(f_\phi(x),c_k'))}.
% \end{equation}

% $s_i$ represents the probability that a sample $x_k$ is predicted as the i-th category. 

% \begin{equation}\label{equ:loss1}
%     s_i=\frac{e^{ a_i}}{\sum_{k}e^{ a_k}}.
%     % p_\phi(y=k|x)=\frac{exp(-d(f_\phi(x),c_k))}{\sum_{k'}exp(-d(f_\phi(x),c_k'))}.
% \end{equation}

% $a_k$ is the k-th element of the softmax output generated by classifier. We train encoder and classifier by minimizing the loss function $L$ that the feature of 3D samples can be extracted and correctly predicted. 

% First, the raw subtomogram data in source domain is utilized, and because of the abundant labeled source data, our encoder and classifier have a perfect prediction performance on source domain. However, the deviation of distribution between target domain and source domain affect their adaptation in target domain adversely. 
 
\subsection{Stage 2: Train the discriminator $\mathcal{D}$}\label{stage2}
After the first training step, the combination of encoders in source domain: $f^\textrm{0} \circ f^\textrm{s}$ and encoders in target domain: $f^\textrm{0} \circ f^\textrm{t}$ and classifier $g$ have a perfect performance in classification of source domain because plentiful labeled source data $X^\textrm{s} = (x^\textrm{s}, y^\textrm{s})$ is supplied. Unfortunately, due to the different experimental imaging parameters in two domains, we can hardly reach satisfactory result in target domain. Thus, the essential part of our proposed method is utilizing unlabeled data and few labeled data in target domain in order to improve its performance in target domain. According to our experiments, even though the amount of labeled data in target domain is scarce, they are notably conductive to the improvement of classification accuracy in test stage.

Inspired by \cite{motiian2017few}, we devise a discriminator $\mathcal{D}$ for Domain Adaptation in the following stages. \cite{motiian2017few} trains the discriminator $\mathcal{D}$ using adversarial training. The method is successful on the popular datasets such as MNIST, USPS and SVHN, because the loss function is easy to design. However, unlike the traditional 2D images, the spatial and structural information of our 3D subtomograms is very complicated and it is severely contaminated by noise. Therefore, it is difficult to train a desirable network using adversarial training because discriminator $\mathcal{D}$ and encoders in source domain: $f^\textrm{0} \circ f^\textrm{s}$ and encoders in target domain: $f^\textrm{0} \circ f^\textrm{t}$ are very hard to converge at the same time and their performance needs to be synchronized. Thus, as much as adversarial training is able to reach a satisfactory result in the traditional image datasets which have relatively high Signal-to-Noise-Ratio(SNR), when it comes to cryo-ET, the drawback of adversarial training would be exposed. Therefore, instead of training discriminator $\mathcal{D}$ and encoders in source domain: $f^\textrm{0} \circ f^\textrm{s}$ and encoders in target domain: $f^\textrm{0} \circ f^\textrm{t}$ alternately like adversarial training, in our model, the discriminator $\mathcal{D}$ is only trained once, and the parameters of the encoders $f^\textrm{s}$, $f^\textrm{t}$, and $f^\textrm{0}$ are not trained during the training of the discriminator $\mathcal{D}$.
%we introduce another way to train discriminator: the parameter of encoder are not updated until we train a desirable discriminator.

In this section, we aim at training a discriminator $\mathcal{D}$ to distinguish the domain of each feature vector, which is described in detailed below.

\subsubsection{Stage 2.1: Preprocessing of encoders in source domain:  $f^\textrm{0} \circ f^\textrm{s}$ and encoders in target domain: $f^\textrm{0} \circ f^\textrm{t}$}\label{stage2_1}

In this stage, we adjust the parameters of encoders in source domain:  $f^\textrm{0} \circ f^\textrm{s}$ and encoders in target domain: $f^\textrm{0} \circ f^\textrm{t}$. By doing this, the encoder $f^\textrm{0} \circ f^\textrm{t}$ is easier to extract the information of subtomograms of target domain. We use a discriminator $\mathcal{D}$ to assist encoders in source domain: $f^\textrm{0} \circ f^\textrm{s}$ and encoders in target domain: $f^\textrm{0} \circ f^\textrm{t}$ to confuse two domains; on the other hand, in order to make the discriminator $\mathcal{D}$ distinguish two domains well, our encoders in source domain: $f^\textrm{0} \circ f^\textrm{s}$ and encoders in target domain: $f^\textrm{0} \circ f^\textrm{t}$ must have the ability to confuse two domains. That's to say,  at the beginning the distributions of feature vectors $T^\textrm{s} := \{ f^\textrm{0} \circ f^\textrm{s}(x^\textrm{s}_i) \}$ and $T^\textrm{t} := \{ f^\textrm{0} \circ f^\textrm{t}(x^\textrm{t}_i) \}$ in the two domains shouldn't have too much notable discrepancy. Otherwise, it would be so easy for the discriminator $\mathcal{D}$ to identify which domain every subtomogram belongs to, and its identification ability can hardly be improved. Therefore, the training of discriminator $\mathcal{D}$ relies on encoders in source domain: $f^\textrm{0} \circ f^\textrm{s}$ and encoders in target domain: $f^\textrm{0} \circ f^\textrm{t}$, and the training of encoders in source domain: $f^\textrm{0} \circ f^\textrm{s}$ and encoders in target domain: $f^\textrm{0} \circ f^\textrm{t}$ relies on discriminator $\mathcal{D}$ too. Unfortunately, neither discriminator $\mathcal{D}$ and encoders in source domain: $f^\textrm{0} \circ f^\textrm{s}$ and encoders in target domain: $f^\textrm{0} \circ f^\textrm{t}$ are fully trained. The encoders $f^\textrm{s}$, $f^\textrm{t}$, and $f^\textrm{0}$ in two domains are all pre-trained on the source data $X^\textrm{s} = (x^\textrm{s}, y^\textrm{s})$, so the parameter of the encoder $f^\textrm{0} \circ f^\textrm{t}$ in target domain is identical to the encoder $f^\textrm{0} \circ f^\textrm{s}$ in source domain. The model trained on data in source domain can hardly extract the feature in target domain very well, which becomes a major obstacle to train a discriminator $\mathcal{D}$.

In order to solve this problem, we use the following tactics. 1) Stage 2.1: Apply unsupervised domain adaptation (UDA) to encoders in source domain: $f^\textrm{0} \circ f^\textrm{s}$ and encoders in target domain: $f^\textrm{0} \circ f^\textrm{t}$. 2) Stage 2.2: Train a discriminator $\mathcal{D}$ with the help of encoders in source domain: $f^\textrm{0} \circ f^\textrm{s}$ and encoders in target domain: $f^\textrm{0} \circ f^\textrm{t}$. 3) Stage 3: Optimize encoders in source domain: $f^\textrm{0} \circ f^\textrm{s}$ and encoders in target domain: $f^\textrm{0} \circ f^\textrm{t}$ with the help of discriminator $\mathcal{D}$. The detailed algorithm of Stage 2.1 is discussed as follows.

% two encoders in source domain: $f^\textrm{0} \circ f^\textrm{s}$ and encoders in target domain: $f^\textrm{0} \circ f^\textrm{t}$ in this stage: one for source data $X^\textrm{s} = (x^\textrm{s}, y^\textrm{s})$ and the other for target $X^\textrm{t} = (x^\textrm{t}, y^\textrm{t})$. 
The encoder $f^\textrm{0} \circ f^\textrm{s}$ trained by source data $X^\textrm{s} = (x^\textrm{s}, y^\textrm{s})$ is pre-trained in the first stage (Equation \ref{equation1}), which we discussed in detail in the Section \ref{encoderandclassifier}. We apply UDA for encoders in source domain: $f^\textrm{0} \circ f^\textrm{s}$ and encoders in target domain: $f^\textrm{0} \circ f^\textrm{t}$ in both domains before training a discriminator $\mathcal{D}$. UDA utilizes unlabeled data in target domain to enable our network the ability to initially confuse the data in two domains.

Specifically, for UDA, inspired by \cite{sun2016deep}, deep correlation alignment (CORAL) is applied to encoders in source domain: $f^\textrm{0} \circ f^\textrm{s}$ and encoders in target domain: $f^\textrm{0} \circ f^\textrm{t}$ to reduce the domain distribution discrepancy between feature vectors $T^\textrm{s}$ and $T^\textrm{t}$ in source and target domains. We implement this method by appending CORAL loss to original classification loss. CORAL loss measures the distribution discrepancy between source domain and target domain in embedding space. We select a set of feature vectors $D^\textrm{s}$ in source domain from $T^\textrm{s}$ and a set of feature vectors $D^\textrm{t}$ in target domain from $T^\textrm{t}$.

Specifically, CORAL loss is defined as:

\begin{equation}
    L^\textrm{CORAL} = \frac{\|C^\textrm{s}-C^\textrm{t}\|^{2}_{F}}{4d^{2}},
\end{equation}
where $\|\cdot\|_{F}$ is the Frobenius norm; $C^\textrm{s}$ is the covariance matrix of $D^\textrm{s}$ and $C^\textrm{t}$ is the covariance matrix of $D^\textrm{t}$; and $d$ is the dimension of the feature vectors $D^\textrm{s}$ and $D^\textrm{t}$. $C^\textrm{s}$ and $C^\textrm{t}$ are calculated by the following equations:
\begin{equation}\label{Cs}
   C^\textrm{s} =  \frac{1}{n^\textrm{s}-1} \bigl[{D^\textrm{s}}^{T} D^\textrm{s}-\frac{1}{n^\textrm{s}}(\textbf{1}^{T} D^\textrm{s})^{T}(\textbf{1}^{T} D^\textrm{s})\bigr],
\end{equation}
\begin{equation}\label{Ct}
  C^\textrm{t} =  \frac{1}{n^\textrm{t}-1} \bigl[{D^\textrm{t}}^{T} D^\textrm{t}-\frac{1}{n^\textrm{t}}(\textbf{1}^{T} D^\textrm{t})^{T}(\textbf{1}^{T} D^\textrm{t})\bigr],
\end{equation}
where \textbf{1} denotes a column vector whose every element is 1; $n^\textrm{s} : =|D^\textrm{s}|$ is number of feature vectors in $D^\textrm{s}$; and $n^\textrm{t} : =|D^\textrm{t}|$ is the number of feature vectors in $D^\textrm{t}$. 

The combined loss is defined as:
\begin{equation}\label{coraltotal}
    L^\textrm{total} = \
 L^\textrm{CORAL} + L^\textrm{C},
\end{equation}
where $L^\textrm{C}$ is the classification loss defined in \ref{equation1_1}.

\begin{figure*}[h!]  
\centering
 \includegraphics[width=0.76\textwidth]{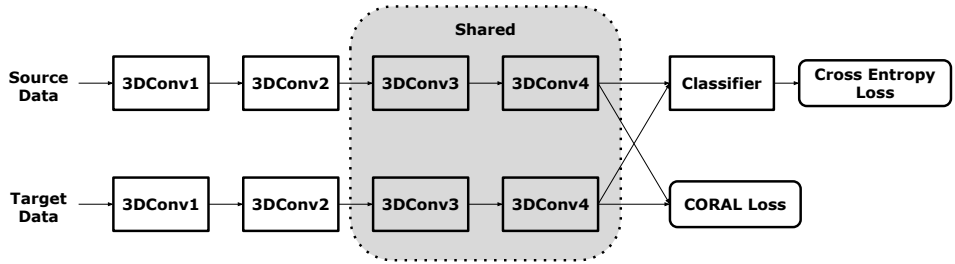}
  \caption{\label{fig:coral_loss} Our model architecture of Unsupervised Domain Adaptation. domain-related layers contain the first and second Convolution Block and domain-independent layers contain the third and last Convolution Block. That's to say, the parameters in the domain-independent layers are shared by data in source domain and target domain.}
      \end{figure*}

The model architecture of UDA is shown in Figure \ref{fig:coral_loss}. Generally, $L^\textrm{CORAL}$ and $L^\textrm{C}$ are opposite: trying to diminish $L^\textrm{CORAL}$ must cause category confusion to encoders in source domain: $f^\textrm{0} \circ f^\textrm{s}$ and encoders in target domain: $f^\textrm{0} \circ f^\textrm{t}$ and classifier $g$ and vice versa. %Therefore, the appropriate ratio between the two loss function is vital. We can calibrate it through hyper-parameter $\alpha$. 
We set $\alpha$ as 500 such that our model can reach a desirable result on target domain.  

We simultaneously input data from two domains, and each batch contains data in both target domain and source domain. We acquire $C^\textrm{s}$ and $C^\textrm{t}$ by calculating the batch covariance \cite{sun2016deep} of subtomograms. In other words, $D^\textrm{s}$ denotes the feature vectors in a subtomogram batch from source domain, and $D^\textrm{t}$ denotes the feature vectors in a subtomogram batch from target domain.

\begin{algorithm}[htb]
\caption{Unsupervised Domain Adaptation Training}\label{algorithm2}
\begin{algorithmic}[1]
\REQUIRE ~~\\
Subtomograms $X^\textrm{s}$ in source domain.\\
Subtomograms $X^\textrm{t}$ in target domain.\\
\ENSURE ~~\\
Trained encoders $f^\textrm{0} \circ f^\textrm{s} $ and $f^\textrm{0} \circ f^\textrm{t}$ and classifier $g$.\\

\STATE \textbf{for} $m$ epochs \textbf{do} \\
\STATE \quad \textbf{for} k steps \textbf{do} \\
% \STATE \qquad Set batch size as \textbf{n} \\
% \STATE \qquad Sample batch $B_s$=${\{(x_{1}^{s},y_{1}^{s}),(x_{2}^{s},y_{2}^{s}),...,(x_{n}^{s},y_{n}^{s})}\}$ from $X_{s}$.  \\
% \STATE \qquad Sample batch $B_t$=$\{(x_{1}^{t},y_{1}^{t}),(x_{2}^{t},y_{2}^{t}),...,(x_{n}^{t},y_{n} ^{t})\}$ from $X_{t}$. \\
% \STATE \qquad Acquire the output of encoder $D_s$ = $f_{0} \circ f_{s}(B_s)$ and $D_t$ = $f_{0} \circ f_{t}(B_t)$.
\STATE \quad Acquire feature vectors batch $D^\textrm{s}$ and $D^\textrm{t}$ from $X^\textrm{s}$ and $X^\textrm{t}$
\STATE \qquad Calculate the covariance matrix $C^\textrm{s}$ and $C^\textrm{t}$ according to equations \ref{Cs} and \ref{Ct}.

\STATE \qquad Update the parameters of encoders $f^\textrm{0} \circ f^\textrm{s} $ and $f^\textrm{0} \circ f^\textrm{t}$ and classifier $g$ by minimizing \ref{coraltotal}.
% the combination of CORAL loss and classification cross entropy loss:

% \begin{eqnarray*}
%     \theta \leftarrow \theta - \frac{1}{n}\beta \nabla_{\theta} [\alpha\frac{\|C_s-C_t\|^{2}_{F}}{4d^{2}}-\mathop{\sum}_{i=1}^{n}y_{i}^{s}log(g\circ f_{\phi}(x_{i}^{s}))]
% \end{eqnarray*}

\STATE \textbf{return} encoders $ f^\textrm{0} \circ f^\textrm{s}$ and $ f^\textrm{0} \circ f^\textrm{t}$ and classifier $g$.

\end{algorithmic}
\end{algorithm}

\subsubsection{Stage 2.2: Update the parameters of discriminator $\mathcal{D}$}\label{SDA}

 In this stage, we aim at training a discriminator $\mathcal{D}$ to differentiate two different domains. In order to fully utilize the label of target domain, inspired by \cite{motiian2017few}, we design the discriminator $\mathcal{D}$ to identify whether two subtomograms are from the same domain and whether they belong to the same category. We consider the condition that the labeled data in target domain is scarce (For example, not more that 7 samples are labeled in each class). These labeled target samples are utilized in this step. We train a discriminator $\mathcal{D}$ in order to distinguish feature vectors $T^\textrm{s}$ and $T^\textrm{t}$ from source domain and target domain with the parameters of encoders in source domain: $f^\textrm{0} \circ f^\textrm{s}$ and encoders in target domain: $f^\textrm{0} \circ f^\textrm{t}$ and classifier $g$ fixed. We combine all of the feature vectors in source domain $T^\textrm{s}$ and labeled feature vectors $T^\textrm{t}_{l}$, and pair feature vectors in $W = T^\textrm{s} \cup T^\textrm{t}_{l}$. There are four kinds of pair combinations: 1) two paired  feature vectors coming from the same domain and category, 2) from the same domain but different categories, 3) from different domains but the same category and 4) from different domains and categories. Therefore, we divide all pairs into 4 groups: $G_{1}$, $G_{2}$, $G_{3}$ and $G_{4}$ to correspond four kinds of pair combination above. The discriminator $\mathcal{D}$ learns to classify each pair into one of the four groups. In each training process, we obtain minibatch by selecting a certain number of feature vector pairs from the 4 groups. The parameters of discriminator $\mathcal{D}$ are updated by the following equation:

\begin{equation}\label{DCD1}
    \theta \leftarrow \theta - \frac{1}{n}\beta \nabla_{\theta} \bigl[-\mathop{\sum}_{i=1}^{n}g_{i}\log(\mathcal{D}\bigl(t_{i}^{1},t_{i}^{2})\bigr)\bigr]
\end{equation}

\noindent where $n$ denotes the size of minibatch. ($t_{i}^{1}$, $t_{i}^{2}$) represents the i-th feature vector pair from minibatch, and $t_{i}^{1}, t_{i}^{2} \in W$. $g_{i} \in \{G_1, G_2, G_3, G_4\}$ represents the group ID of the i-th pair of minibatch. We use the function $\mathcal{D}(\cdot)$ to denote the discriminator $\mathcal{D}$.

% $G_{1}$ contains pairs each of whose items belongs to the same category and domain; $G_{2}$ contains pairs each of whose items belongs to the same category but different domains; $G_{3}$ contains pairs each of whose items belongs to different categories but the same domain; $G_{4}$ contains pairs each of whose items belongs to different categories and domains. 

\begin{figure*}[h!]
\centering
 \includegraphics[width=1.0\textwidth]{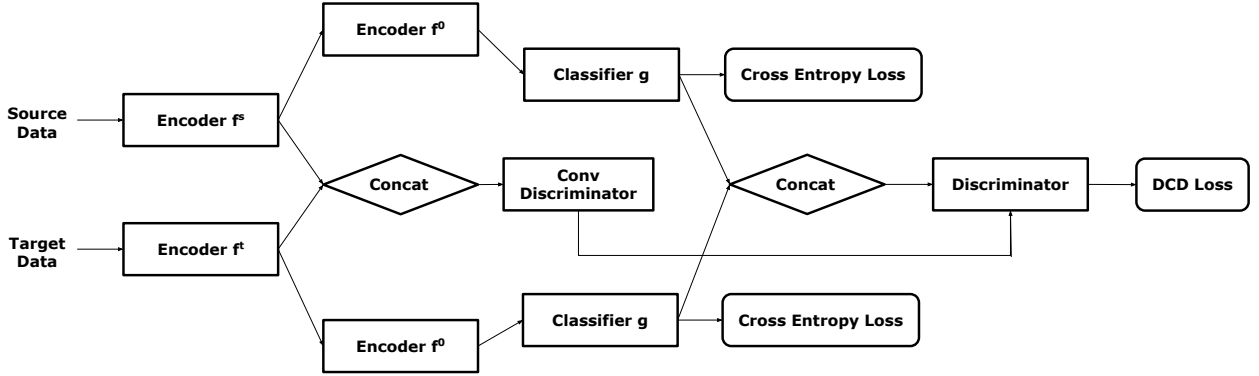}
  \caption{\label{fig:dcd_loss} Our model architecture of Supervised Domain Adaptation (Stages 2.2 and 3). }  
\end{figure*}

The architecture of discriminator is showed in Figure \ref{fig:dcd_loss}. The discriminator $\mathcal{D}$ contains the 3D discriminator and 1D discriminator corresponding to our partly-shared encoder $f_{\phi}$. The output of domain-independent layers (feature vectors $T^\textrm{s}, T^\textrm{t}$) and output of domain-related layers are both discriminated, because we assume that the input distribution of  domain-independent layers has low correlation with domain variation. 3D discriminator distinguishes output domain of domain-related layers. 1D discriminator integrates the output of 3D discriminator and feature vectors $T^\textrm{s}, T^\textrm{t}$ then calculates the group ID of each pairs.

% \begin{figure*}[h!]
% \centering
%  \includegraphics[width=1.0\textwidth]{DCD_Loss.eps}}
%   \caption{\label{fig:dcd_loss} Our model architecture of Supervised Domain Adaptation (Stages 2.2 and 3). }  
% \end{figure*} 
      
\subsection{Stage 3: Fine-tune the encoder $f_{\phi}$}\label{stage3}

After training the discriminator $\mathcal{D}$, we fine-tune the encoders in source domain: $f^\textrm{0} \circ f^\textrm{s}$ and encoders in target domain: $f^\textrm{0} \circ f^\textrm{t}$ and classifier $g$ again with the parameters of discriminator $\mathcal{D}$ frozen. We need to make discriminator $\mathcal{D}$ confused between $G_{1}$ and $G_{2}$, and also between $G_{3}$ and $G_{4}$ by updating the parameters of encoders in source domain: $f^\textrm{0} \circ f^\textrm{s}$ and encoders in target domain: $f^\textrm{0} \circ f^\textrm{t}$, which is measured by the domain-class discriminator (DCD) loss \cite{motiian2017few}:
\begin{equation}
    L^{\textrm{DCD}} = -E\bigl[y_{G_{1}}\log\bigl(D(G_{2})\bigr)-y_{G_{3}}\log\bigl(D(G_{4})\bigr)\bigr],
\end{equation}
where $y_{G_{i}}$ represents the ID of $G_{i}$. Therefore the total loss can be denoted as:
\begin{equation}\label{totalDCD}
    L^{\textrm{total}} = \gamma L^{\textrm{DCD}} + L^\textrm{s} + L^\textrm{t},
\end{equation}
where $L^\textrm{s}$ and $L^\textrm{t}$ are the cross entropy loss functions to the classification of source domain and target domain.

\begin{algorithm}[htb]
\caption{Supervised Domain Adaptation Training}
\begin{algorithmic}[1]
\REQUIRE ~~\\
Cryo-ET data in source domain: $X^\textrm{s}$.\\
Labeled cryo-ET data in target domain: $X^\textrm{t}$.\\
\ENSURE ~~\\
Trained encoders in source domain: $f^\textrm{0} \circ f^\textrm{s}$ and encoders in target domain: $f^\textrm{0} \circ f^\textrm{t}$, classifier $g$ and discriminator $\mathcal{D}$.\\
\STATE Sample groups $G_1$,$G_2$,$G_3$ and $G_4$ \\
\STATE \textbf{for} $m$ epochs \textbf{do} \\
% \STATE \quad \textbf{for} k steps \textbf{do} \\
% \STATE \qquad Set minibatch size as \textbf{n} \\
% \STATE \qquad Sample minibatch from $G_1$,$G_2$,$G_3$ and $G_4$.
% \STATE \qquad Sample minibatch $B_s$=$\{(x_{1}^{s},x_{1}^{t},y_{1}^{s},y_{1}^{t},g_{1}),(x_{2}^{s},x_{2}^{t},y_{2}^{s},y_{2}^{t},g_{2}),$...,\\ $(x_{n}^{s},x_{n}^{t},y_{n}^{s},y_{n}^{t},g_{n})\}$ among $G_1$,$G_2$,$G_3$ and $G_4$. \\
\STATE \quad Update $\mathcal{D}$
 with encoders in source domain: $f^\textrm{0} \circ f^\textrm{s}$ and encoders in target domain: $f^\textrm{0} \circ f^\textrm{t}$ and classifier $g$ fixed by minimizing \ref{DCD1}.

\STATE \textbf{for} $m$ epochs \textbf{do} \\
% \STATE \quad \textbf{for} k steps \textbf{do} \\
% \STATE \qquad Set minibatch size as \textbf{n} \\
% \STATE \qquad Sample minibatch $B_s$=$\{(x_{1}^{s},x_{1}^{t},y_{1}^{s},y_{1}^{t},g_{1}),(x_{2}^{s},x_{2}^{t},y_{2}^{s},y_{2}^{t},g_{2}),$...,\\ $(x_{n}^{s},x_{n}^{t},y_{n}^{s},y_{n}^{t},g_{n})\}$ among $G_1$,$G_2$,$G_3$ and $G_4$. \\
\STATE \quad Update encoders in source domain: $f^\textrm{0} \circ f^\textrm{s}$ and encoders in target domain: $f^\textrm{0} \circ f^\textrm{t}$ and classifier $g$ with discriminator $\mathcal{D}$ fixed by minimizing \ref{totalDCD}.
\STATE \textbf{return} encoders $f^\textrm{0} \circ f^\textrm{s}$ and $ f^\textrm{0} \circ f^\textrm{t}$, classifier $g$ and discriminator $\mathcal{D}$
% \STATE \qquad Acquire the output of encoder $D_s$ = $f_{0} \circ f_{s}(B_s)$ and $D_t$ = $f_{0} \circ f_{t}(B_t)$.
% \STATE \qquad Calculate the covariance matrix of $D_{s}$ and $D_{t}$ according to equations \ref{Cs} and \ref{Ct}.

% \STATE \qquad Update the parameters $\theta$ of encoder and classification by minimizing the combination of CORAL loss and classification cross entropy loss:

% \begin{eqnarray*}
%     \theta \leftarrow \theta - \frac{1}{n}\beta \nabla_{\theta} [\frac{\|C_s-C_t\|^{2}_{F}}{4d^{2}}-\mathop{\sum}_{i=1}^{n}y_{i}^{s}log(g\circ f_{\phi}(x_{i}^{s}))]
% \end{eqnarray*}

% \STATE \textbf{return} encoder $f_{\phi}$ and classifier $g$.

\end{algorithmic}
\end{algorithm}

\section{Results}
\subsection{Datasets}

\subsubsection{Simulated Subtomograms}\label{simulatedsubtomogram}
The simulated subtomograms of $35^{3}$ voxels are generated similar to \cite{xu2017deep}. Two simulated subtomogram dataset batches $S_1$, $S_2$ are provided to realize the domain adaptation process. $S_1$ is acquired through 2.2mm spherical aberration, -10$\mu   m$ defocus and 300kV voltage. $S_2$ is acquired through 2mm spherical aberration, -5$\mu   m$ defocus and 300kV voltage. Each dataset batch contains four datasets with different SNR levels (0.03, 0.05, 0.1, 0.5, 1000). Specifically, there are 43 macromolecular classes in each dataset. All of macromolecular classes are collected from PDB2VOL program \cite{abola1984protein}, and each class in each dataset contains 100 subtomograms. 

% Some examples of simulated subtomogram have been presented in \ref{fig:demo}.

\subsubsection{Real Subtomogram Datasets}
We test our model on two real subtomogram datasets $S_1$ and $S_2$.
$S_1$ is extracted from rat neuron tomograms \cite{guo2018situ}, containing Membrane, Ribosome, TRiC, Single Capped Proteasome, Double Capped Proteasome and NULL class(the subtomogram with no macromolecule). Its SNR is 0.01, and the tilt angle ranges from $-50^{\circ}$ to $+70^{\circ}$.

$S_2$ is a single particle dataset from EMPIAR \cite{noble2018reducing}, containing Rabbit Muscle Aldolase, Glutamate Dehydrogenase, DNAB Helicase-helicase, T20S Proteasome, Apoferritin, Hemagglutinin and Insulin-bound Insulin Receptor. Its SNR is 0.5, with tilt angle range $-60^{\circ}$ to $+60^{\circ}$, size $28^3$ voxels, and voxel spacing 0.94nm.

\begin{table}[!h]
\centering
\caption{\label{results_simulated1}The classification accuracy of the dataset from target domain. The result in each cell represents the accuracy of CORAL \cite{sun2016deep}, Sliced Wasserstein Distance\cite{gabourie2020system}, finetune, FADA and our method from top to bottom. The highest accuracy in each cell is highlighted. It shows that the prediction accuracy of our method surpasses the baseline methods in most of the cases. }
\begin{tabular}{|c|c|c|c|c|c|c|}
    \hline
    ~ & \multicolumn{6}{|c|}{Target Domain} \\
    \hline
    \multirow{13}*{Source Domain} & SNR & 1000 & 0.5 & 0.1 & 0.05 & 0.03 \\
    \cline{2-7}
    \multirow{13}*{~} & \multirow{5}*{1000} & 0.470 & 0.066 & 0.049 & 0.034 & 0.024 \\
    \multirow{13}*{~} & \multirow{5}*{~} & 0.442 & 0.148 & 0.095 & 0.078 & 0.063\\
    \multirow{13}*{~} & \multirow{5}*{~} & 0.513 & 0.211 & 0.083 & 0.062 & 0.050\\
    \multirow{13}*{~} & \multirow{5}*{~} & 0.664 & 0.404 & 0.185 & 0.161 & 0.146\\
    \multirow{13}*{~} & \multirow{5}*{~} & \textbf{0.761} & \textbf{0.518} & \textbf{0.253} & \textbf{0.196} & \textbf{0.177}\\

    \cline{2-7}
    \multirow{13}*{~} & \multirow{5}*{0.5} & 0.189 & 0.369 & 0.219 & 0.125 & 0.104\\
    \multirow{13}*{~} & \multirow{5}*{~} & 0.321 & 0.374 & 0.244 & 0.144 & 0.150 \\
    \multirow{13}*{~} & \multirow{5}*{~} & 0.387 & 0.416 & 0.204 & 0.137 & 0.111 \\
    \multirow{13}*{~} & \multirow{5}*{~} & \textbf{0.660} & \textbf{0.577} & 0.328 & \textbf{0.255} & 0.171\\
    \multirow{13}*{~} & \multirow{5}*{~} & 0.601 & 0.532 & \textbf{0.332} & 0.254 & \textbf{0.203}\\

    \cline{2-7}
    \multirow{13}*{~} & \multirow{5}*{0.1} & 0.107 & 0.24 & 0.237 & 0.188 & 0.166 \\
    \multirow{13}*{~} & \multirow{5}*{~} & 0.125 & 0.250 & 0.230 & 0.166 & 0.143 \\
    \multirow{13}*{~} & \multirow{5}*{~} & 0.280 & 0.285 & 0.263 & 0.184 & 0.147 \\
    \multirow{13}*{~} & \multirow{5}*{~} & 0.415 & 0.436 & 0.297 & 0.231 & 0.170\\
    \multirow{13}*{~} & \multirow{5}*{~} & \textbf{0.513} & \textbf{0.456} & \textbf{0.332} & \textbf{0.257} & \textbf{0.218}\\

    \cline{2-7}
    \multirow{13}*{~} & \multirow{5}*{0.05} & 0.034 & 0.147 & 0.197 & 0.145 & 0.13 \\
    \multirow{13}*{~} & \multirow{5}*{~} & 0.057 & 0.170 & 0.126 & 0.152 & 0.150 \\
    \multirow{13}*{~} & \multirow{5}*{~} & 0.184 & 0.238 & 0.203 & 0.191 & 0.137 \\
    \multirow{13}*{~} & \multirow{5}*{~} & 0.280 & 0.292 & 0.231 & 0.205 & 0.176\\
    \multirow{13}*{~} & \multirow{5}*{~} & \textbf{0.439} & \textbf{0.374} & \textbf{0.292} & \textbf{0.256} & \textbf{0.235}\\

    \cline{2-7}
    \multirow{13}*{~} & \multirow{5}*{0.03} & 0.045 & 0.117 & 0.115 & 0.122 & 0.127 \\
    \multirow{13}*{~} & \multirow{5}*{~} & 0.061 & 0.123 & 0.098 & 0.088 & 0.106 \\
    \multirow{13}*{~} & \multirow{5}*{~} & 0.089 & 0.190 & 0.166 & 0.166 & 0.148 \\
    \multirow{13}*{~} & \multirow{5}*{~} & \textbf{0.276} & 0.229 & 0.202 & 0.194 & 0.177\\
    \multirow{13}*{~} & \multirow{5}*{~} & 0.218 & \textbf{0.243} & \textbf{0.211} & \textbf{0.200} & \textbf{0.200}\\
    \hline
    \end{tabular}

\end{table}

% \begin{figure}[!htb]
% \centering 
% \includegraphics[width=0.2\textwidth]{demo.eps}
% \end{figure} 
\subsection{Classification Results}
We conduct experiments respectively with finetune, FADA and our methods on simulated datasets and real datasets, and compare the results of these methods. Finally, we demonstrate the superiority of our method on the simulated and real datasets.

\subsubsection{Results of Simulated Datasets}\label{sec321}

In this experiment, $A_s$ is denoted as source domain and $A_t$ is denoted as target domain. For facilitating computation, we randomly sample 100 subtomograms from each class. Table \ref{results_simulated1} presents the prediction accuracy in these methods.

\subsubsection{Results of Cross-domain Prediction of Real Subtomograms}\label{sec323}
 The real datasets are acquired in the very complicated environment, causing the heterogeneity of subtomograms and very low SNR comparing to simulated dataset. This characteristic of experimantal datasets poses a challenge to the macromolecule classification. 

\begin{figure*}[h!]  
\centering
 \includegraphics[width=0.8\textwidth]{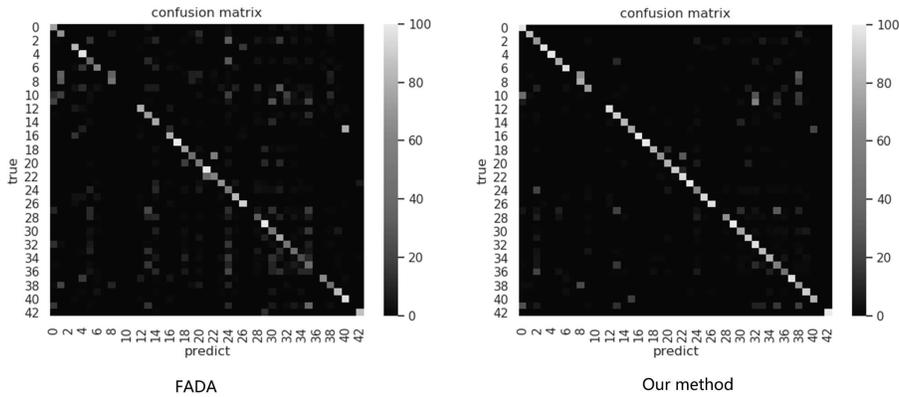}
  \caption{\label{fig:confusion} The confusion matrix in our method and baseline method. In view that FADA is far more better than other baselines, we compare the confusion matrix in our method to those in FADA. The left is the confusion matrix of FADA, the right is the confusion matrix of our method.}
  \end{figure*}

Five simulated datasets in $A_{s}$ and $A_{t}$ with different SNR(1000, 0.5, 0.1, 0.05, 0.03) are utilized. Each of the simulated dataset acts as source domain, and their classes are the same as target domain. and two real subtomogram datasets are acted as the target domain. Table \ref{results_experimental} shows the classification results on all of the methods. The result in each cell represents the prediction accuracy in real dataset, and the confusion matrices have been showed in \ref{fig:confusion}. Additionally, 3 and 7 labeled samples are selected in target domain for supervised training in FADA and our method.

\begin{table}[!h] 
\centering
\caption{\label{results_experimental}The classification accuracy on the real dataset. The row of the cell denotes which method is utilized and the column of the cell denotes the SNR of source domain. The first row is FADA and the second row is our method. It is obvious that the superiority of our method increases as the SNR becomes lower comparing to FADA.}

   \begin{tabular}{|c|c|c|c|c|c|c|c|}
    \hline     
    \multicolumn{3}{|c|}{\multirow{2}{*}{~}} & \multicolumn{5}{|c|}{Target Domain} \\
    \cline{4-8}
    \multicolumn{3}{|c|}{} & 1000 & 0.5 & 0.1 & 0.05 & 0.03 \\
    \hline

    \multirow{8}*{Source Domain} & \multirow{4}*{3 shot} & \multirow{2}*{$S_1$} & \textbf{0.801} & 0.626 & 0.453 & 0.535 & 0.538\\
    \multirow{8}*{~} & \multirow{4}*{~} & \multirow{2}*{~} & 0.732 & \textbf{0.720} & \textbf{0.608} & \textbf{0.606} & \textbf{0.586}\\

    \cline{3-8}
    \multirow{8}*{~} & \multirow{4}*{~} & \multirow{2}*{$S_2$} & 0.705 & 0.731 & 0.793 & 0.748 & 0.655\\
    \multirow{8}*{~} & \multirow{4}*{~} & \multirow{2}*{~} & \textbf{0.788} & \textbf{0.733} & \textbf{0.891} & \textbf{0.849} & \textbf{0.774}\\

    \cline{2-8}
    \multirow{8}*{} & \multirow{4}*{7 shot} & \multirow{2}*{$S_1$} & \textbf{0.842} & 0.664 & 0.760 & 0.679 & 0.690\\
    \multirow{8}*{~} & \multirow{4}*{~} & \multirow{2}*{~} & 0.774 & \textbf{0.805} & \textbf{0.791} & \textbf{0.719} & \textbf{0.701}\\

    \cline{3-8}
    \multirow{8}*{~} & \multirow{4}*{~} & \multirow{2}*{$S_2$} & \textbf{0.959} & 0.952 & 0.947 & 0.953 & 0.796\\
    \multirow{8}*{~} & \multirow{4}*{~} & \multirow{2}*{~} &  0.833 & \textbf{0.969} & \textbf{0.971} & \textbf{0.954} & \textbf{0.958}\\
    \hline

    \end{tabular}
\end{table}  

% \begin{figure}[h!]  
%  \includegraphics[width=0.5\textwidth]{demo.eps}
%   \caption{\label{fig:demo} Some examples of our subtomogram density map.}  
%       \end{figure} 

\section{Conclusion}

Recently, Cryo-Electron Tomography emerges as a powerful tool for systematic \textit{in situ} visualization of the structural and spatial information of macromolecules in single cells. However, due to high structural complexity and the imaging limits, the classification of subtomograms is very difficult. Supervised deep learning has become the most powerful method for large scale subtomogram classification. However, the construction of high quality training data is laborious. In such case it is beneficial to utilize another already annotated dataset to train neural network model. However, there often exists a systematic image intensity distribution difference between the annotated dataset and target dataset. In such case the model trained on another annotated dataset may have a poor performance in target domain.
%Macromolecule classification is critical for understanding macromolecules and their interactions with other subcellular components.
In this paper, we propose a Few-shot Domain Adaptation method to for cross-domain subtomogram classification. Our method combines Unsupervised Domain Adaptation and  Supervised Domain Adaptation: we first train a discriminator $\mathcal{D}$ to identify the domain of each subtomogram, and we utilize the discriminator $\mathcal{D}$ to assist us the process of SDA. To the best of our knowledge, this is the first work to apply semi-supervised Domain Adaptation on subtomogram classification. We conduct experiments on simulated dataset and real dataset, and the prediction accuracy of our methods surpasses the baseline methods. Therefore, our method can be effectively applied to the subtomogram classification from a new domain with only a  few labeled samples supplied. Our work represents an important step toward fully utilizing deep learning for subtomogram classification, which is critical for the large-scale and systematic \textit{in situ} recognition and recovery of macromolecular structures in single cells captured by cryo-ET.

% \section*{Acknowledgements}

% Text Text Text Text Text Text  Text Text.  \cite{Boffelli03} might want to know about  text
% text text text\vspace*{-12pt}

\section*{Funding}
This work was supported in part by U.S. National Institutes of Health (NIH) grant P41GM103712 and R01GM134020, U.S. National Science Foundation (NSF) grant DBI-1949629 and IIS-2007595, and Mark Foundation for Cancer Research grant 19-044-ASP. XZ was supported by a fellowship from Carnegie Mellon University's Center for Machine Learning and Health. 

\appendix
\section{Few Shot Domain Adaptation for \textit{in situ} Macromolecule Structural Classification in Cryo-electron Tomograms -- Supplementary Document}
\subsection{Related work}
Current analysis of cryo-ET includes template matching \cite{beck2009visual}. First we create the templates for every class; and for every subtomogram, we calculate the matching score between the template and itself. The method is straight-forward and easy to realize, but the computation complexity is unbearable, especially on the data which has countless classes and more dimensions compared to traditional images. What's more, because of the intense disruption of noise, the error rate of this method is very high. 

Another method utilizes unsupervised subtomogram classification (e.g. \cite{xu2012high}). Now there are a set of subtomograms which correspond to k classes. First we initialize k class centers, each of which represents the average of subtomograms in each class, and therefore all of subtomograms can be classified by computing the distance to k class centers. Second, after labeling all of subtomograms, we redirect k class centers by calculating average of labeled subtomograms in each class. By computing the two above steps iteratively, we can approximately obtain the label of every subtomogram. This method doesn't require the label of any subtomogram, reducing the workload of labeling our data. Nevertheless, even if some tactics eliminating noises has been elaborated in this paper, the noises are still remaining a severe problems in our subtomogram classification, which extremely affect the performance in this method adversely.  

There is another straightforward resolution that we can implement transfer learning into subtomogram classification. We generate simulated dataset in the computer as source domain and set real dataset as target domain. With the development of Neural Network, many Deep Learning models use this tactic to solve this problem. \cite{moebel:tel-02153877} proposed an unsupervised classification method with transfer learning. First we train a CNN model by simulated dataset. In the second stage, we remove the last layer from the original model and then extract the feature vector of the real dataset. In the end, we apply k-means clustering to the feature vector of the real dataset.

Recently \cite{lin2019adversarial} applies Unsupervised Domain Adaptation to subtomogram classification, in order to resolve the situation, in which source domain(train dataset) and target domain(test dataset) have different image intensity distribution. Even though it reaches a desirable performance in target domain, there still remains limitations because no label in target domain is utilized; what's more, adversarial training is used in this paper. In Section 2.2, we have discussed in detail that adversarial training is hard to be convergent when using Cryo-ET. Comparing to this paper, we success to utilize the label information in target domain and further improve its performance. 

There are two mainstream ways for current Unsupervised Domain Adaptation methods to decrease image intensity distribution. First, the training dataset is used to optimize our model and later the parameter of this model is fine-tuned by test dataset. Even if the label information of test dataset isn't available, some of its global features, such as mean and covariance, can still be calculated. This kind of information is crucial for us to fine-tune the parameters of our model. For instance, \cite{sun2016deep} and \cite{tzeng2017adversarial} use this way as domain adaptation. The second way is transforming the data in target domain, making its distribution more similar to the data in source domain. Compared to the first way, parameters of the model would not be fine-tuned. For instance, \cite{alam2018speaker} use this way as domain adaptation. In this paper, whitening and re-coloring, which utilizes the covariance of data in source domain and data in target domain, are applied to data in source domain. The source data being transformed are used to train the classification model. Because transformed source data has the similar distribution with target data, the model can reach a desirable result on target domain.

Compared to Unsupervised Domain Adaptation, Few-shot Domain Adaptation utilizes the whole data in source domain and very few labeled data in target domain. The core idea is very similar to Few Shot Learning\cite{snell2017prototypical}: we require our model to learn the features in very few images. However, the two fields still have very significant difference. Few shot learning needs to learn the image features whose labels aren't presented in training dataset, while few shot domain adaptation needs to learn the image features whose domain are different from training dataset. 

\subsection{Time complexity}

We test the time complexity of FSFT and other Deep Learning methods, which is presented in Table \ref{app4}.

\begin{table}[!htb] 
\centering
\caption{\label{app4} This table lists cost time of five Deep Learning method. FSFT costs less time than SWD and Fine-tune while costs more time than FADA and CORAL.}

   \begin{tabular}{|c|c|}
    \hline     
    Model & Time Cost(s) \\
    \hline
    CORAL & 323.66 \\
    \hline
    SWD & 1797.37 \\
    \hline
    Fine-tune & 1002.08 \\
    \hline
    FADA & 554.80 \\
    \hline
    FSFT & 921.36 \\
    \hline
    \end{tabular}
\end{table}  

From the table, CORAL, as the simplest method, costs the least time. Compared to FADA, FSFT add Deep CORAL as one of crucial stage, and its model is more complex. These changes introduce more computation, in order to have a better performance in subtomogram classification.

\subsection{Result Analysis}
We conduct some experiments to analyze to verify the effectiveness of FSFT. We generate 23 classes for the simulated datasets $S_1$ and $S_2$ which are mentioned in Section 3.1.1. In this section, we want to verify the effectiveness of each stage and each contribution we proposed. All the experiments in this section are conducted on these datasets.

Firstly, in the task of subtomogram classification, we split the whole training procedure into 3 stages which are tightly linked. We verify that each stage plays an important role in improving the classification precision. Table \ref{app1} presents the improvement of each stage in FSFT. Unsupervised Domain Adaptation is used in Stage 2 and Supervised Domain Adaptation is used in Stage 3. The combination of them enable encoders $f^\textrm{0} \circ f^\textrm{t}$ to adapt the target domain. 

\begin{table}[!htb] 
\centering
\caption{\label{app1}We calculate the classification accuracy in each stage. This table shows that in Stage 2.1, Deep CORAL method improves the accuracy from 37.2\% to 70.9\%; In Stage 2.2, we only update the parameter of discriminator, so the prediction accuracy is the same as Stage 2.1. In Stage 3, Fine-tune the encoder $f_{\phi}$ improves the accuracy from 70.9\% to 95.3\%}

   \begin{tabular}{|c|c|}
    \hline     
    Stage & Accuracy \\
    \hline
    Stage 1 & 37.2\% \\
    \hline
    Stage 2.1 & 70.9\% \\
    \hline
    Stage 2.2 & 70.9\% \\
    \hline
    Stage 3 & \textbf{95.3\%} \\
    \hline
    
    \end{tabular}
\end{table}  

Secondly, the contributions mentioned in Section 1 are effective in improving the performance of FSFT. In order to verify their effectiveness, we remove each contribution in FSFT and test its performance in target domain. In table \ref{app2}, FSFT method we proposed realizes the best performance, while others can't reach the optimal accuracy compared to FSFT.

\begin{table}[!htb] 
\centering
\caption{\label{app2}Accuracy of ablation study. Row 1 corresponds to FSFT we proposed; Row 2 corresponds to FSFT without Stage 2.1; Row 3 corresponds to FSFT which use GAN to train discriminator and encoder; Row 4 corresponds to FSFT which only use 1D discriminator.}

   \begin{tabular}{|c|c|}
    \hline     
    Model & Accuracy \\
    \hline
    FSFT & \textbf{95.3\%} \\
    \hline
    FSFT without CORAL & 79.1\% \\
    \hline
    FSFT with GAN & 94.2\% \\
    \hline
    FSFT without 3D Discriminator & 95.1\% \\
    \hline
    
    \end{tabular}
\end{table} 

\begin{figure}[h]  
\centering
 \includegraphics[width=0.6\textwidth]{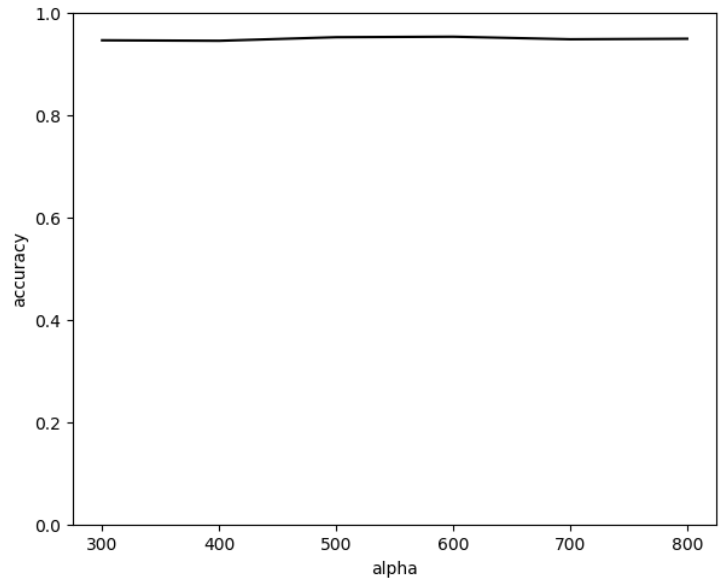}
  \caption{\label{fig:alpha} This picture shows how $\alpha$ affect the prediction accuracy.  }  
\end{figure} 

%\FloatBarrier

\subsection{Hyper-parameter Adjustment}
In this section, we discuss how to choose the hyper-parameter, for example, $\alpha$ in equation 6. In figure \ref{fig:alpha}, the best result is 0.954\%. Accuracy, as $\alpha$ changes, the accuracy nearly stays constant. The value of $\alpha$ will have little effect on the performance of our model.

\bibliographystyle{unsrt}
\bibliography{main}

\end{document}